Long Paper*

# Development of IoT Smart Greenhouse System for Hydroponic Gardens

Arcel Christian H. Austria
Computer Engineering Department, LPU-Cavite, Philippines
arcelaustria12@gmail.com
(corresponding author)

John Simon Fabros
Computer Engineering Department, LPU-Cavite, Philippines
johnsimonfabros@gmail.com
(corresponding author)

Kurt Russel G. Sumilang
Computer Engineering Department, LPU-Cavite, Philippines
kurt.sumilang@lpunetwork.edu.ph
(corresponding author)

Jocelyn Bernardino
Computer Engineering Department, LPU-Cavite, Philippines
jocelyn.bernardino@lpu.edu.ph
(corresponding author)

Anabella C. Doctor
Computer Engineering Department, LPU-Cavite, Philippines
anabella.doctor@lpu.edu.ph
(corresponding author)

*Date received*: January 29, 2023
*Date received in revised form*: March 8, 2023; March 11, 2023
*Date accepted*: March 26, 2023

Recommended citation:

Austria, A.C. H., Fabros, J.S., Sumilang, K.R. G., Bernardino, J., & Doctor, A.C. (2023). Development of IoT Smart Greenhouse System for Hydroponic Gardens. *International Journal of Computing Sciences Research*, 7, 2111-2136. https://doi.org/10.25147/ijcsr.2017.001.1.149

**Special Issue on International Research Conference on Computer Engineering and Technology Education 2023 (IRCCETE 2023). Guest Associate Editors:* **Dr. Nelson C. Rodelas, PCpE** (*Computer Engineering Department, College of Engineering, University of the East-*

*Caloocan City; nelson.rodelas@ue.edu.ph*) and **Engr. Ana Antoniette C. Illahi**, PCpE *(Asst. Professor & Vice Chair, Department of Computer and Electronics Engineering, Gokongwei College of Engineering, De La Salle University, Taft Ave., Manila; ana.illahi@dlsu.edu.ph*).

**Abstract**

*Purpose* – This study focused on the development of a smart greenhouse system for hydroponic gardens with the adaptation of the Internet of Things and monitored through mobile as one of the solutions towards the negative effects of the worlds booming population, never ending - shrinking of arable lands, and the effect of climate change drastically in our environments where innovative techniques are highly encouraged to assure long-term agricultural, food security, and sustainability.

*Method* – To achieve the goal of the study, the researchers employed applied research. The researchers created an actual hydroponic greenhouse system with completely developing plants, and automation in examining and monitoring the water pH level, light, water, and greenhouse temperature, as well as humidity which is linked to ThingSpeak. Alpha testing is performed by the respondents to confirm that the developed system is ready for beta testing as well evaluation process.

*Results* – The developed SMART Greenhouse monitoring system was tested and evaluated to confirm its reliability, functions, and usability under ISO 9126 evaluation criteria. The respondents who include casual plant owners and experts in hydroponic gardening able to test and evaluate the prototype, and the mobile application to monitor the parameters with the results of 7.77 for pH level, 83 for light, 27.94°C for water temperature, 27°C for greenhouse temperature, and 75% for humidity with a descriptive result in both software and hardware as "Very Good" with a mean average of 4.06 which means that the developed technology is useful and recommended.

*Conclusion* – The SMART Greenhouse System for Hydroponic Garden is used as an alternative tool, solution, and innovation technique towards food shortages due to climate change, land shortages, and low farming environments. With the application of the Internet of Things in a greenhouse, automated monitoring of plants becomes convenient and easier.

*Recommendations* – The proponents highly suggest the use of solar energy for the pump power, prototype wiring should be improved, the usage of a high-end model of Arduino to address more sensors and devices for a larger arsenal of data collected, enclosures of the device to ensure safety, and mobile application updates such as bug fixes and have an e-manual of the whole systems.

## INTRODUCTION

The fast change in climate, every nation's increasing population, and the never-ending shrinking of arable lands now necessitate innovative techniques to assure long-term agricultural and food security (Velazquez et al., 2022). Dait (2022) stated that climate change is one of the most alarming occurrences worldwide, with a severe impact mainly in the Philippines because of insufficient capability to adapt to global warming that will cause lower agricultural production. In the Philippine economy, agriculture is very important. It employs around 40% of Filipino employees and provides an average of 20% of the country's GDP. This production is mostly derived from agribusiness, which accounts for over 70% of total agricultural output. Crop cultivation is the most important agricultural activity. But even so, the value of agricultural production fell by 2.5 percent from January to September 2021. Accordingly, a crisis in Philippine agriculture has resulted from a significant drop in productivity, inefficiency, high production costs, and insufficient government support for the industry, among other factors, and crop cultivation, designed as an alternative way of solving these types of problems, also comes with shortcomings (Briones, 2021).

As a result, greenhouse agriculture is viewed as a very viable alternative and long-term solution to future food shortages because it allows farmers and users to regulate their local environment and grow crops all year long, even in extreme weather conditions (Rayhana et al., 2020). As cited in the study (Acharya et al., 2021), the most important crops with a high potential for improving farmers' income due to the growing demand are vegetables, which can be planted in all possible places and ways. Accordingly, a large number of vegetables can be easily grown in a hydroponics system. Hydroponics is a way of doing soil-free gardening and is known as a new method of farming, as stated by (Patel et al., 2018). The hydroponic system encourages plant growth by using water instead of soil, and the features that the environment provides are extremely important for the plant to develop properly.

Greenhouse farming is now one of the world's fastest-growing businesses, a diverse production of crops in any season using a house-like structure made of glass or plastic materials where the roof is usually covered with transparent material to maintain the necessary climatic conditions for plant development (such as temperature, humidity, illumination, and so on), as well as to protect the plants from pests, illness, and bad environmental circumstances. As noted in the study (Ortner & Agren, 2019), plant upkeep must be prioritized where pH level, oxygen, sunshine exposure, and water level must be all considered to provide a healthy and productive environment for the plant. Concerning the said statement, a monitoring system with the application of a microcontroller like Arduino along with several operational and management issues, the integration and adaptation of Internet of Things technology will be a great tool and could make a difference.

The ever-evolving Internet of Things (IoT) technologies, to which certain people have very common knowledge, include smart sensors and devices, network topologies that can be applied in varieties of situations, large data analytics, as well as it being intelligence, and reliability when it comes to decision-making stated in the study of (Lakshmanan et al., 2020). These are some of the few advancements of IoT which are thought to be the answer to the key challenges facing greenhouse farming, such as crop management, water, pH level, and nutrient sufficiency, as well as climate control within the greenhouse. Furthermore, with this being implemented widely, it may be employed in a variety of situations to help farmers, croppers, and planters grow their businesses particularly if it is managed with wireless communication based. It is stated that wired systems that manage greenhouse farms tend to be more complex and harder than using wireless communication which is now being used in smart agriculture to replace the cable system, which was difficult to install and run.

The goal of this study aims to: (a) design and develop an IoT-Based controlling and monitoring apparatus on a greenhouse for hydroponic gardens, (b) design and develop a controlling and monitoring android mobile application for the greenhouse system using MIT App Inventor, (c) construct an IoT-Based Greenhouse Hydroponic system that can monitor the pH level, light, water and greenhouse temperature that include humidity, (d) to develop an IoT-Based Greenhouse Hydroponic system that manipulates pH level, and to lower the water temperature, and (e) test and evaluate the Greenhouse System using ISO 9126, in terms of its functionality, reliability, and usability.

## Conceptual Framework

Figure 1 shows the conceptual framework of the study. The conceptual framework includes the possible Input, Process, and Output of the research study. The input will focus more on the initial investigation of the study which includes the knowledge, hardware, and software requirements of the study. The process part of the framework is about the development of the system from planning up to summarizing it. Lastly, the output should be the objective of the study.

## METHODOLOGY

### *Research Design*

The proponents of this study employed applied research to quantify and grasp enough knowledge about the issues and problems that individuals experienced when cultivating plants in a climate-dependent setting (Kothari, 2008). The data gathered was utilized by the proponents to analyze statistically and methodically with the use of the quantifiable data that has been taken from varieties of tests within the study. With this, understanding the theories that support and facilitate the study will allow researchers to visualize the difficulties with existing farming systems and how they may be remedied by developing an IoT - based Smart Greenhouse System for Hydroponic Farming. In terms of completely developing plants in this greenhouse system, the researchers will be examined and monitored the benefits and drawbacks of having this automated system linked to ThingSpeak. The hydroponic greenhouse is in a six by twelve feet plot of land and has a height of eight feet on the right side and six feet on the left side. The researchers wanted to discover what components and characteristics will be suitable and most appropriate to have a more reliable agricultural system by employing the applicable strategy where a thorough investigation was carried out to see how items altered to provide and aid the data collected.

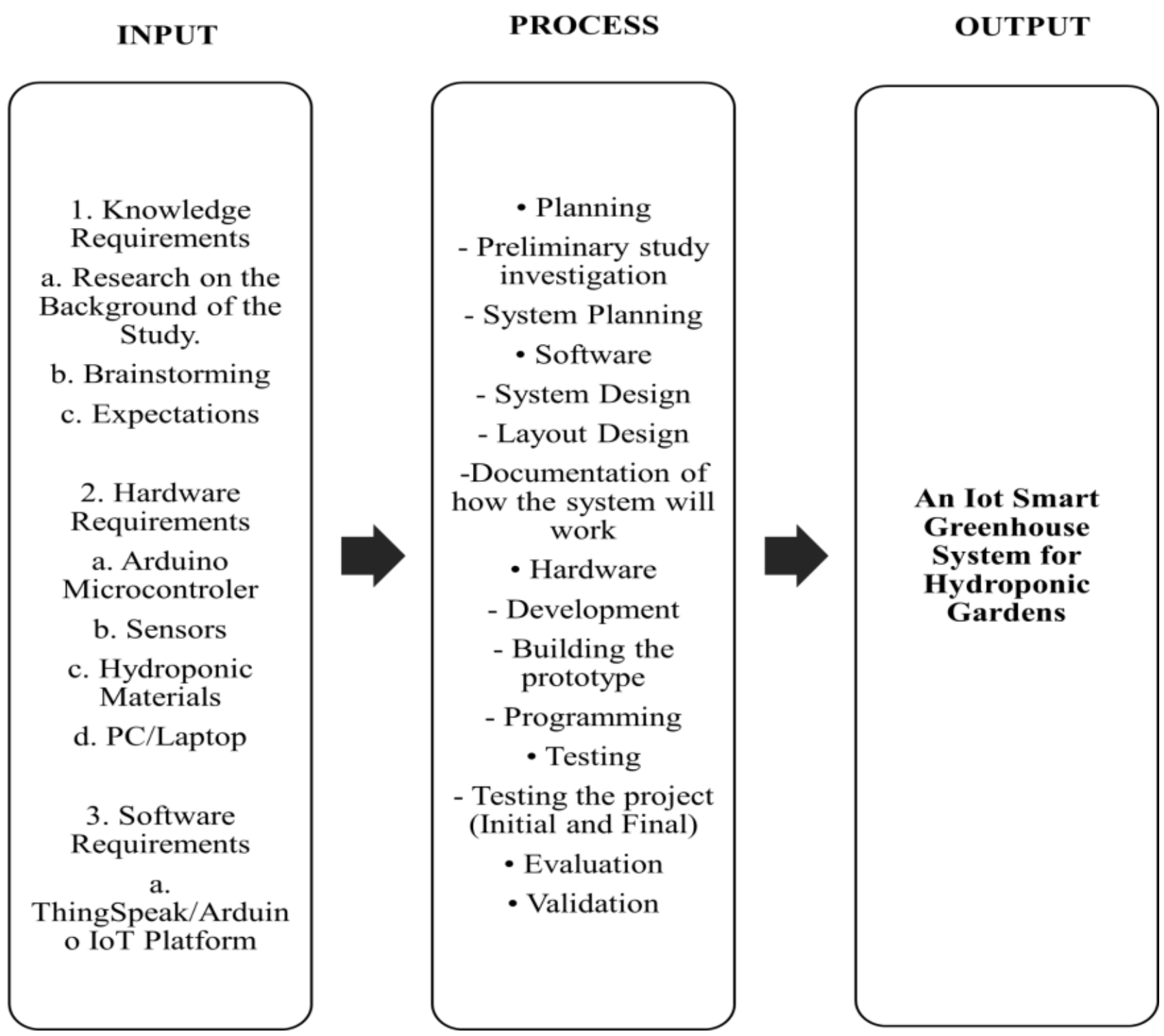

*Figure 1.* Conceptual Framework of the Study

## *Data Gathering Procedure*

The proponents conducted the data-gathering procedures by means of observing and taking note of the changes and results from which the product itself has been shown. The said data vary from the pH level of the water, its light, temperature, as well as oxygen levels, and finally by testing if the structure and support of the greenhouse farm itself hold its weight. The results of these specific variables were gathered through a specific sensor of the said variable, with one being appointed to each component the proponents aim to test within their product. In this way, through the help of ThingSpeak, the researchers recorded the reliable data displayed which is shown in the later part of the document and then furthered explained thoroughly.

***Project Construction, System Model, Schematic Diagram, Control System, and Sensors Flowchart***

To construct the outcome of the study, the proponents strictly follow the identified project design model and diagrams. Figure 2 shows the overall flow on how the project was constructed form the beginning up to the end where the outcome of the study obtained.

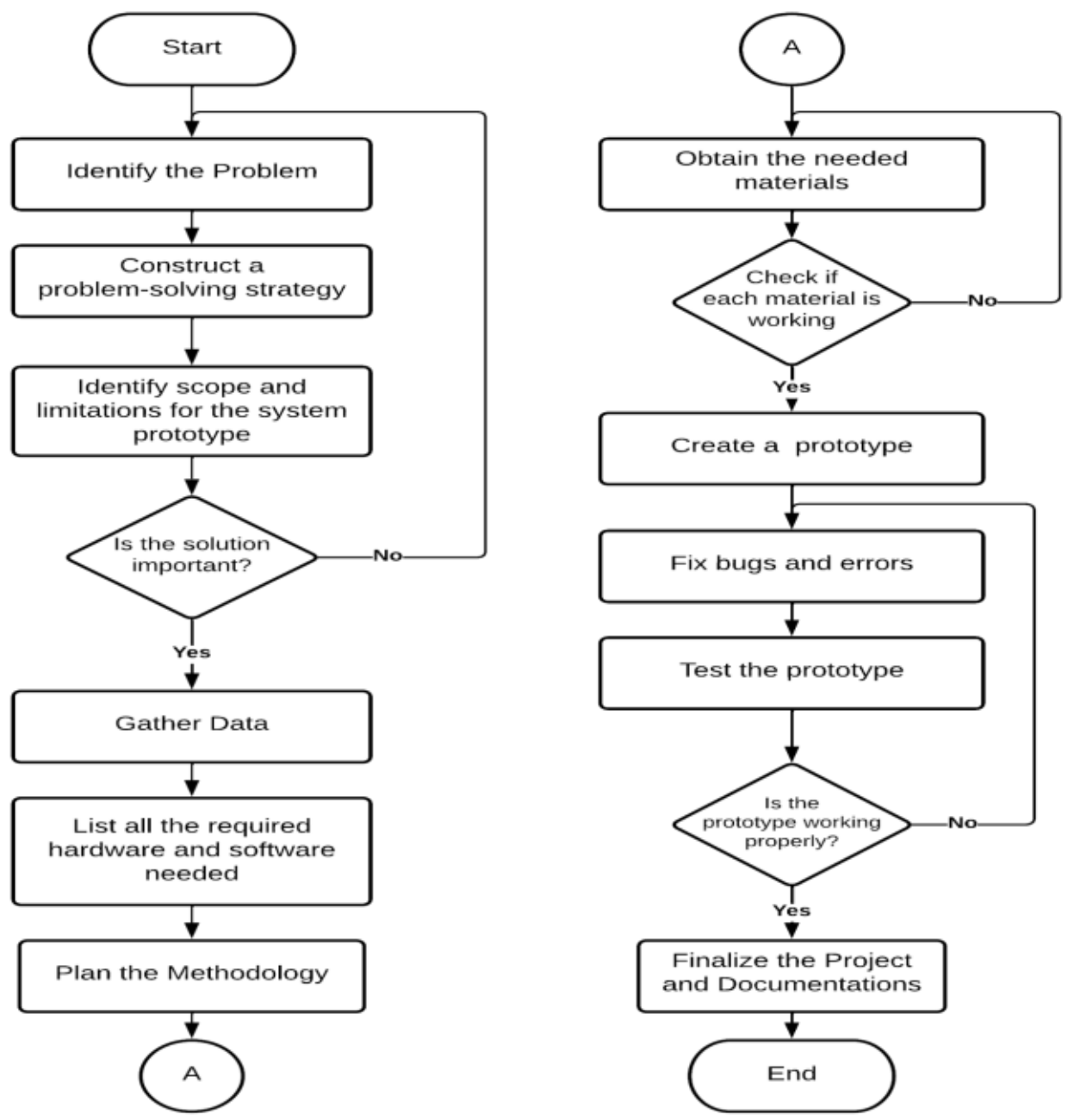

*Figure* 2. Project Construction Flowchart

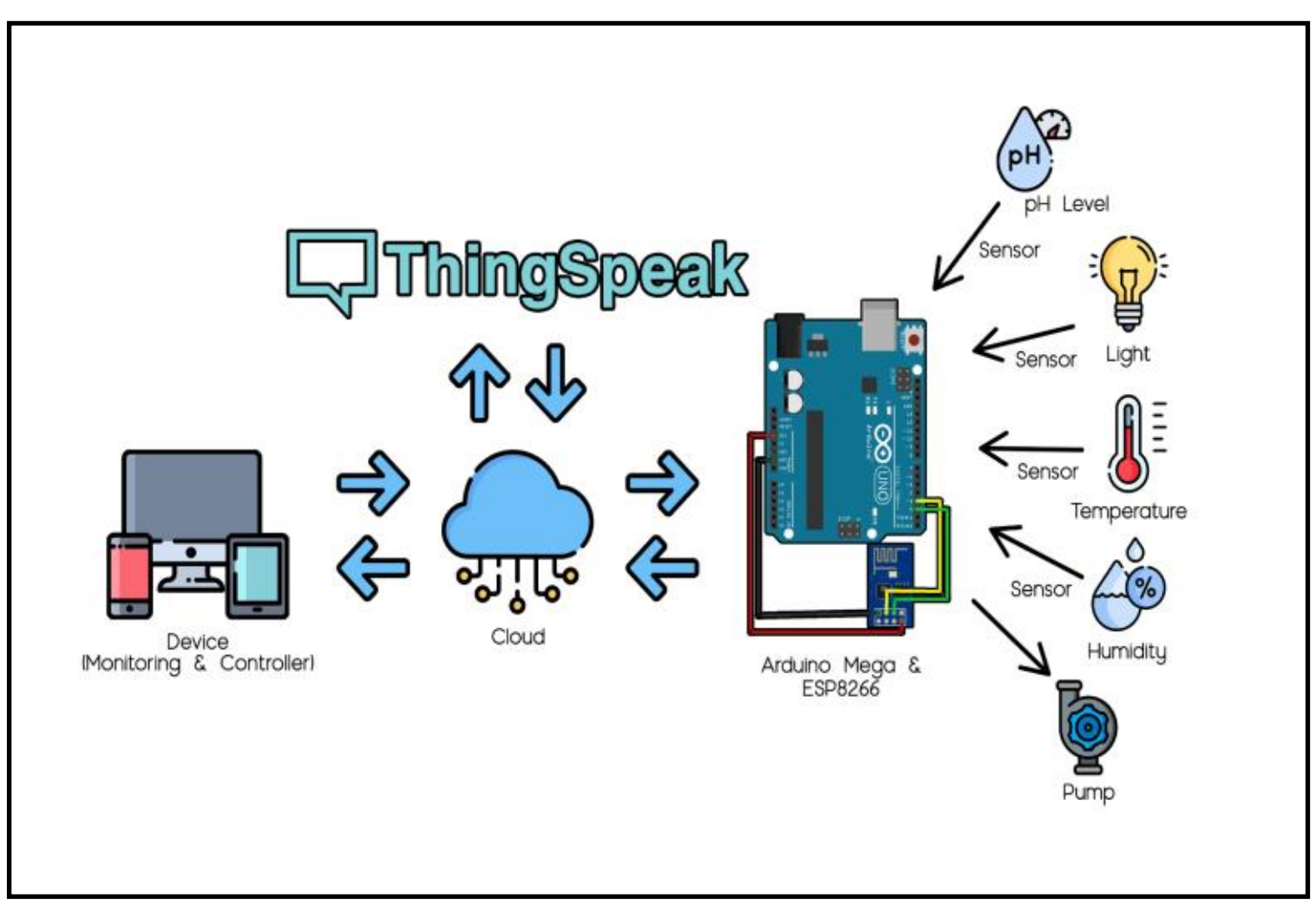

*Figure* 3. System Model

Figure 3 shows the system model for the IoT Smart Greenhouse System on how it works automatically. The device is connected through the cloud with the utilization of ThingSpeak which then associates itself with the Arduino Mega that checks for the pH level of the water, the plant's oxygen, light, temperature, and humidity levels with the use of its specific sensors.

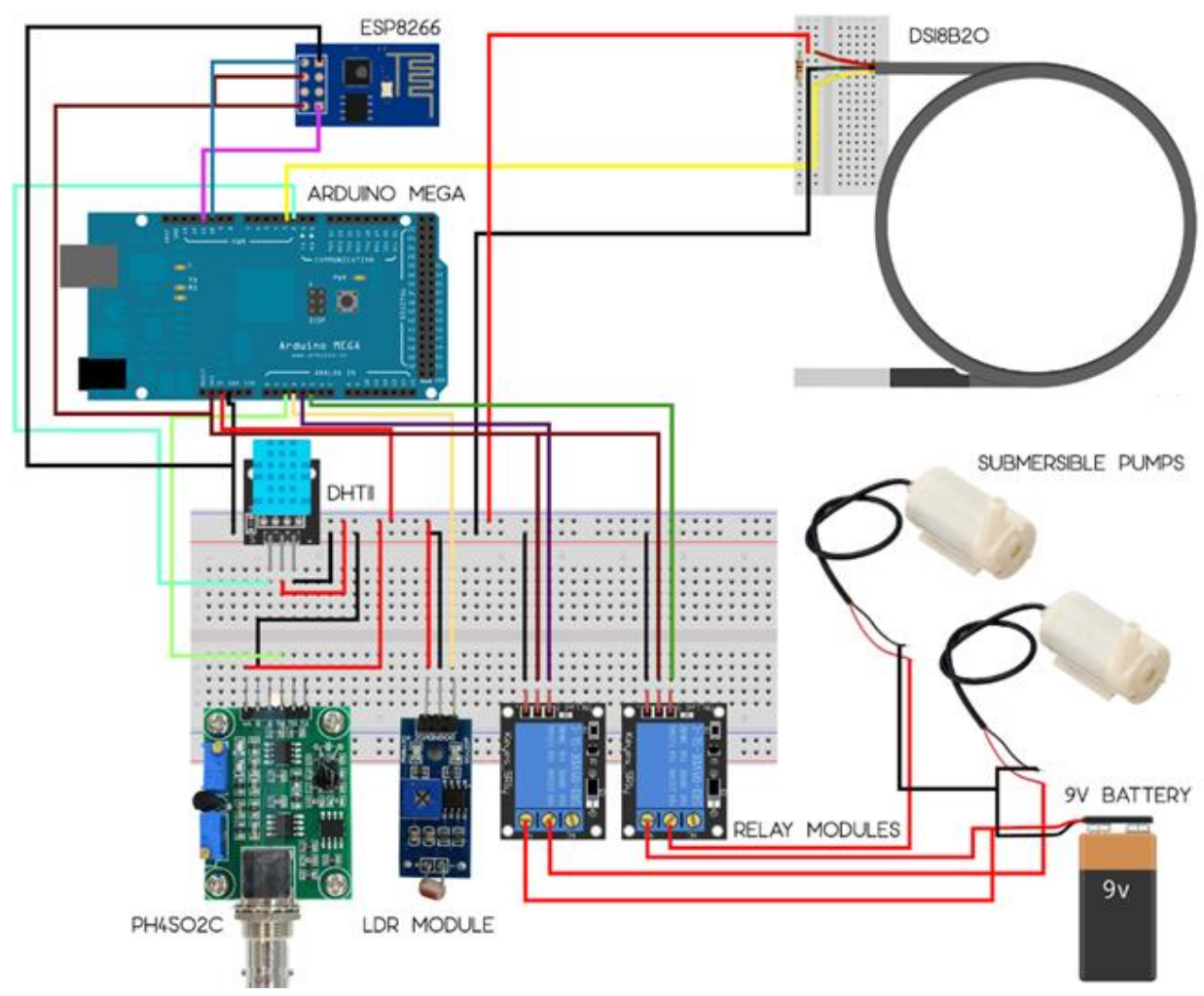

*Figure 4*. Schematic Diagram of the System

Figure 4 shows the schematic diagram of the system. It includes the connection of 4 sensors to the Arduino Mega: DHT11, LDR Module, DS18B20, and PH4502C. It also shows the connection of the Relay Module and the Pump.

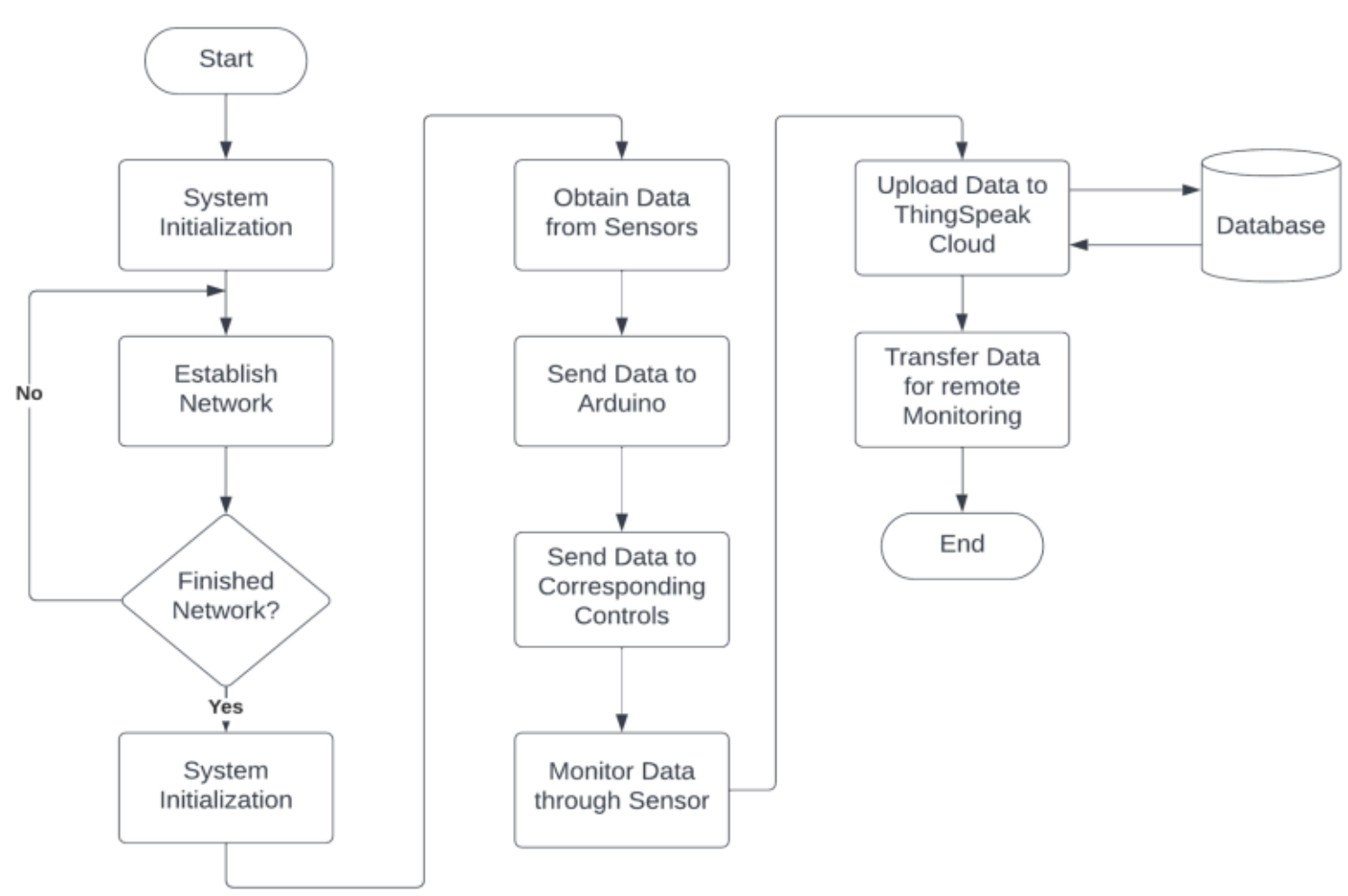

*Figure 5*. Control System Flowchart

The control system manages the threshold of the system. It controls the quality of the outcome of the system and ensures that it follows what it intended to do. Figure 5 shows the flow of the control system for the IoT Smart Greenhouse System for Hydroponic Farm.

## *Project Testing and Evaluation Procedures*

The main purpose of this stage is to ensure that the developed project meets the expected functionality, reliability, and usability of the system as required. Santelices (2013) stated that alpha and beta testing are used to validate and verify system meets the technical requirements that guided its design and development. The test includes a step–by–step procedure on how the hardware components is interconnected with each other thru circuits and how the developed software integrated into the hardware systems and works as one system. Testing and operation procedures are performed and shown to the evaluators during the evaluation to ensure that all features of the system function and performed according to the required specifications of the system.

To validate the outcome of this study, the proponents conducted two processes such as: (1) the testing results from the prototype conducted by the proponents were compared to the results coming from commercialized devices, and (2) through the test survey where the proponents picked a total of ten (10) respondents to assess the prototype's functionality, dependability, and usability, basing on the convenience sampling technique since this number is proven sufficient to construct a foundation of solid evaluation findings (Graglia, 2022). The survey contains questions about the functionality, usability, and reliability of the prototype. The ISO 9126 survey was used as an adaptation for the system's questionnaire since the results reflect the acceptability of the system in terms of being functional, dependable, and usable. Some questions adapted from MARS, the mobile application rating system, and a few questions created by the proponents were used to create the mobile application questionnaire. For the mobile application, the proponents will be using MARS (Mobile Application Rating Scale) for its criteria.

The survey employs a five-point Likert scale, as recommended by comparable surveys, as part of the convenience sample method (Pollfish, 2021).

Table 1. Likert's Scale

| Scale | Equivalent | Mean Rating Score |
|---|---|---|
| 5 | Strongly Agree | 4.20 to 5.00 |
| 4 | Agree | 3.40 to 4.19 |
| 3 | Slightly Agree | 2.60 to 3.39 |
| 2 | Disagree | 1.80 to 2.59 |
| 1 | Strongly Disagree | 1.00 to 1.79 |

The Likert scale shown in Table 1 was used in the project evaluation, its equivalent interpretation and mean score rating which was designed to capture ranges probability after averaging the scores (Doctor & Benito, 2019). Furthermore, this study used descriptive statistics in the interpretation and analysis of the data collected. This includes the frequencies and weighted means. A simple frequency count was applied in tallying responses while a weighted mean was utilized to determine the average response for every criteria description. The weighted average mean for each of the three criteria was also computed by adding all the weighted means divided by the total number of descriptions for every criterion. Lastly, an overall weighted mean was calculated to get the overall quality characteristics of the system perceived by their respondents.

## RESULTS AND DISCUSSION

### *Working SMART Greenhouse in a Small Farm*

This study designed a working greenhouse for a small hydroponic farm. The size of the structure shown in Figure 6 is 6 ft high on one side and 8 ft on the other side, 12 ft long, and 6 ft wide. For the hydroponic system of the greenhouse, the materials used are 10 feet of the black PVC pipe 3 inches in diameter, for the plants, and 10 feet of blue PVC pipe half an inch in diameter for the connection of the black pipes and to the source.

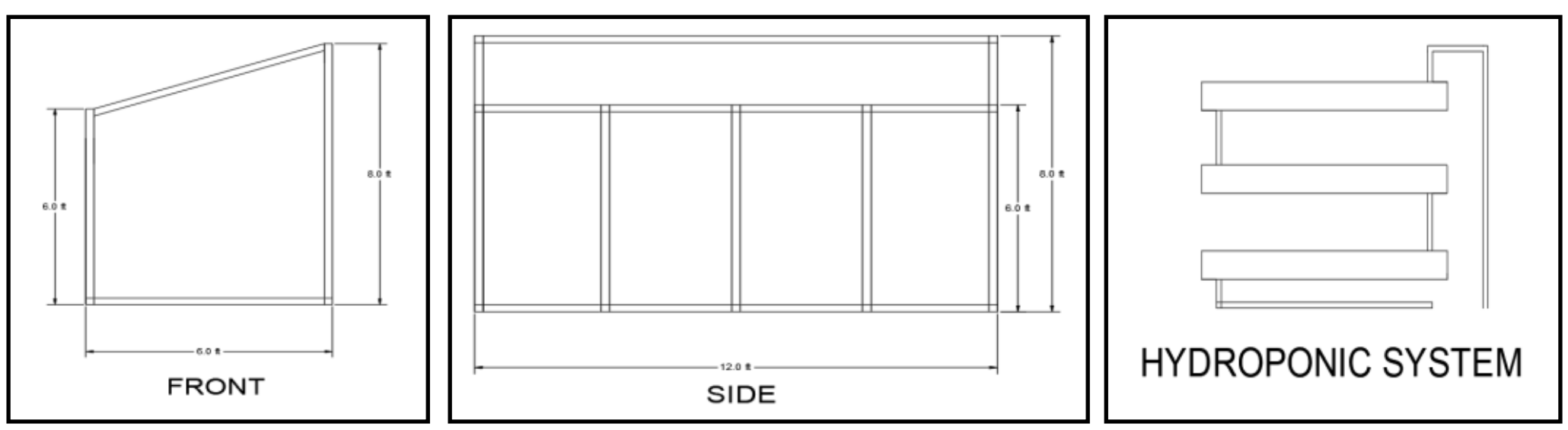

*Figure* 6. Greenhouse Layout Structure

Furthermore, the whole system is composed of Arduino Mega which is placed at the center of the system that acts as the brain, where the sensors, relay, pumps, and ESP8266 are connected to control the system automatically.

Figure 7 shows the actual presentation of the whole system which is located inside the greenhouse. It shows the front and side view of the greenhouse, and the systems (hardware components) are located beside the drum of the main pump.

Figure 8-9 shows the detailed major hardware components of the project. The ESP8266 is a module that sends the gathered parameters to the cloud of ThingSpeak and is connected and positioned at the back of the Arduino Mega. The Light sensor is positioned below the highest pipe, and it measures the light level in the greenhouse. The placement of water temperature sensor is inside the drum of the main pump which senses the water temperature of the hydroponic garden.

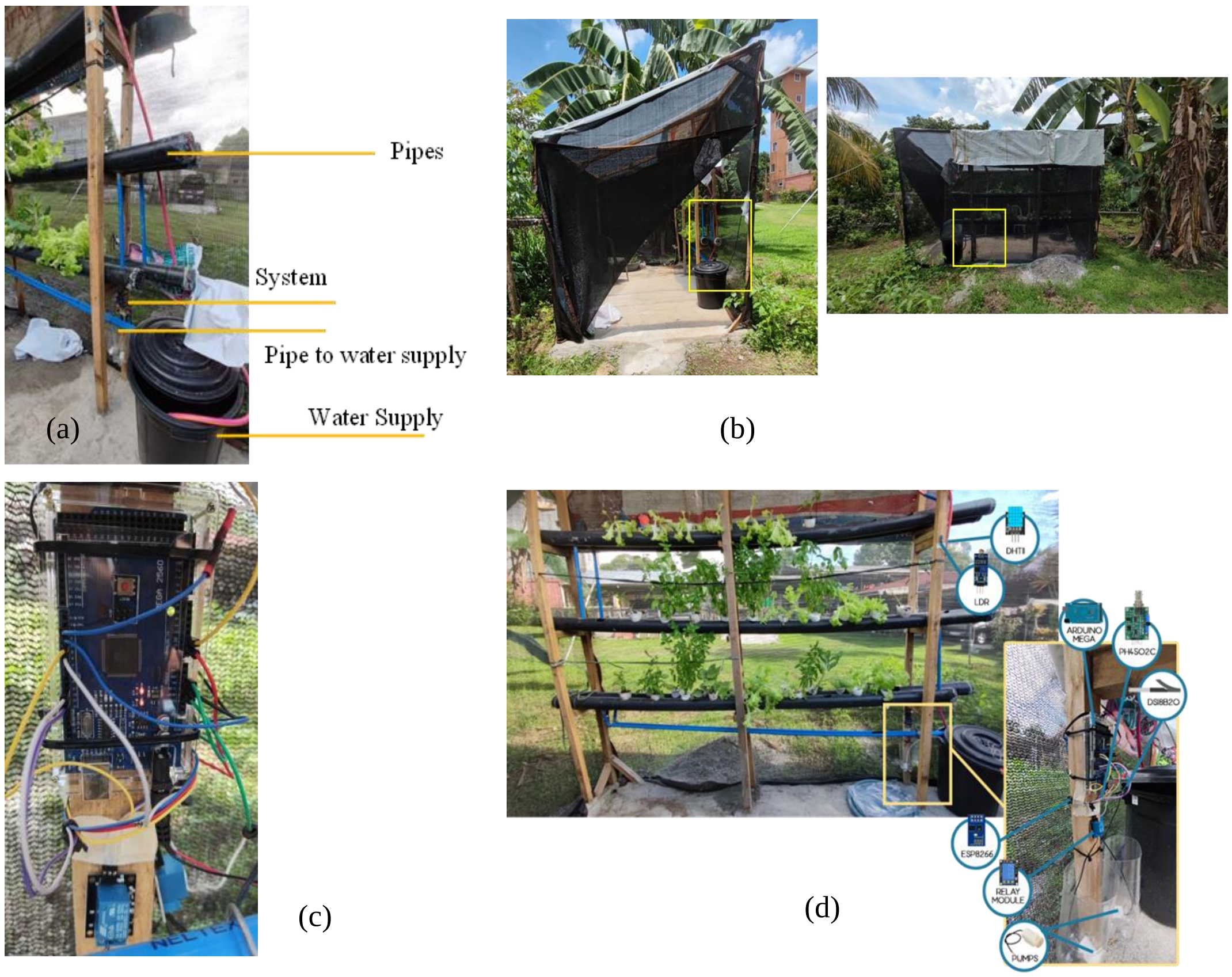

*Figure 7*. Actual Demonstration Model of SMART Greenhouse with Hydroponic Garden

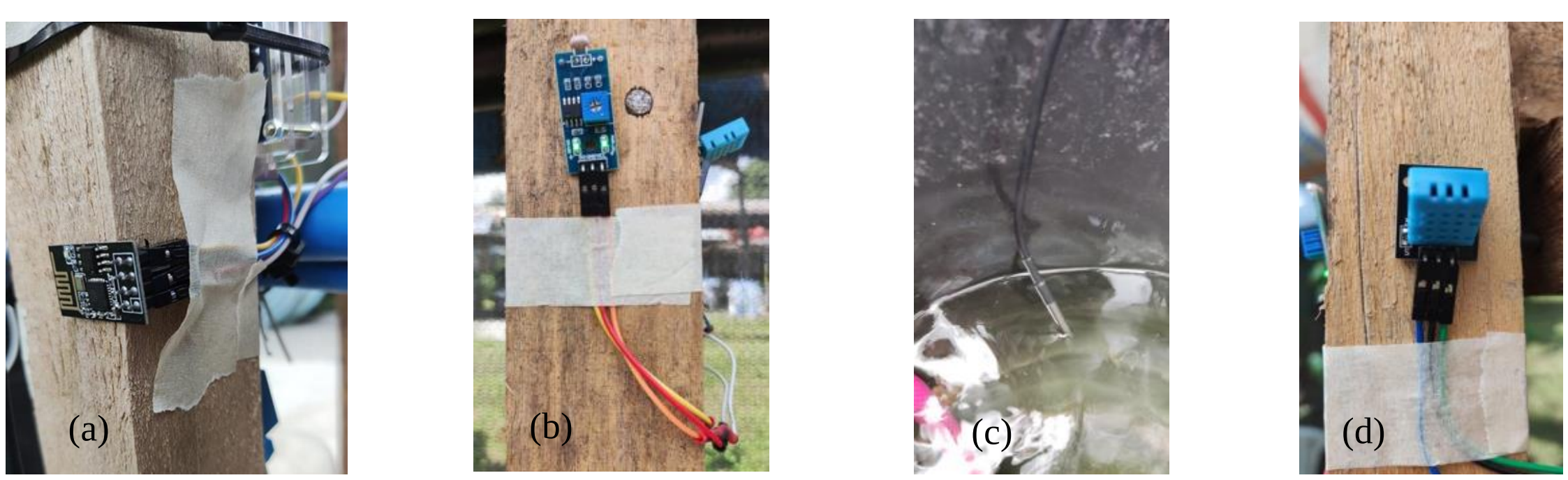

*Figure 8*. (a) ESP8266, (b) Light Sensor Photoresistor, (c) Water Temperature, and (d) DHT-22 Humidity Temperature.

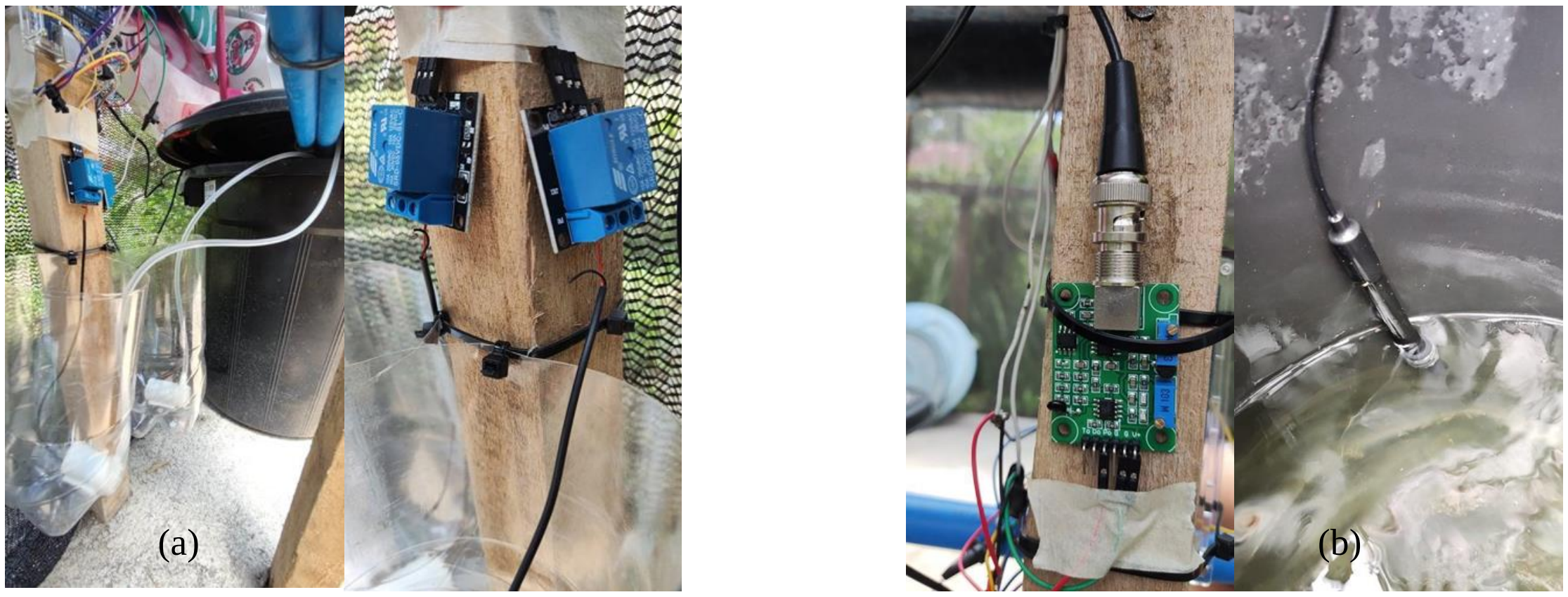

*Figure 9*. (a) Relay Modules with the Submersible Pumps attached, and
(b) pH4502c with the pH probe attached

The DHTT-22 humidity sensor is positioned the same as the light sensor to measure the temperature and humidity of the greenhouse. Moreover, the placement of the pH sensor with the pH probe attached is inside the drum of the main pump which senses the pH level of the water in the hydroponic garden. The Relay modules are connected to the Arduino which can be automatically triggered when the temperature and pH level senses a certain amount level and the submersible pumps will be turned on automatically together with the Relay module.

### *Controlling and Monitoring Android Mobile Application of the SMART Greenhouse*

To achieve the required function of the system in terms of control and monitoring of the SMART greenhouse, ThingSpeak program is needed. ThingSpeak is a cloud-based open-source software/site Internet of Things analytics tool that allows developers to gather, visualize, and analyze live data streams.

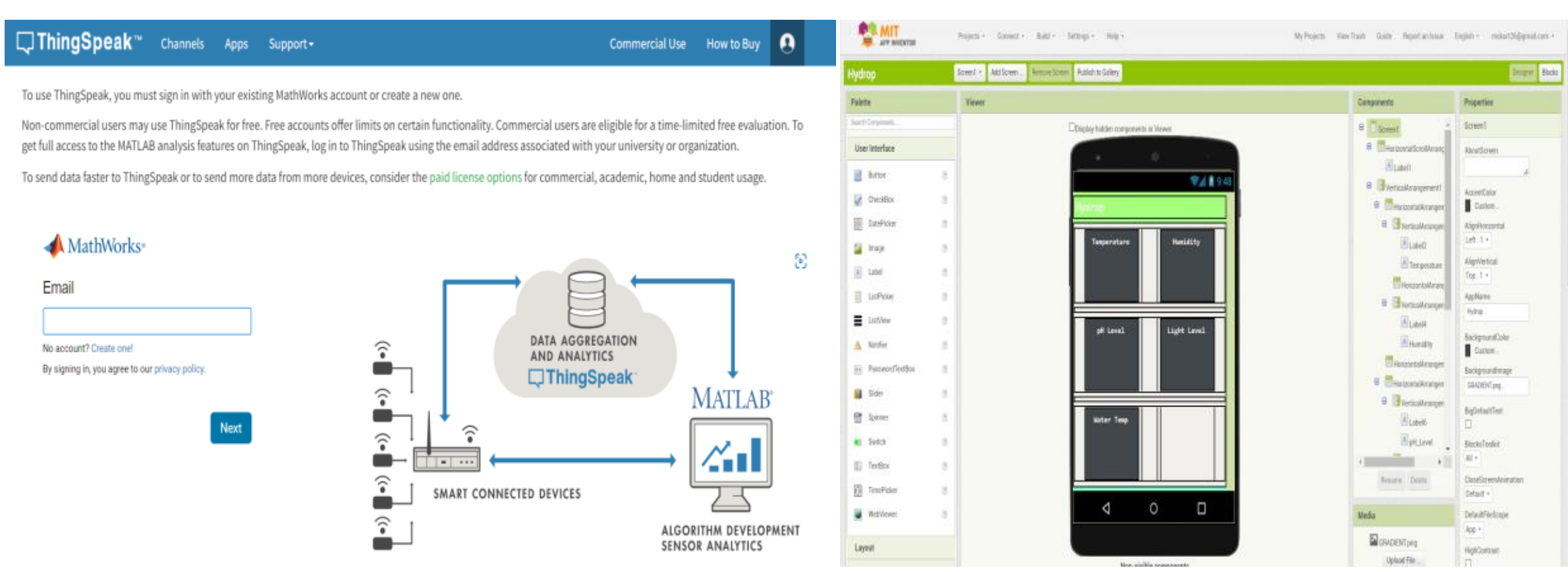

*Figure 10.a*. Thingspeak Login Page  *Figure 10.b*. MIT App Inventor Designer

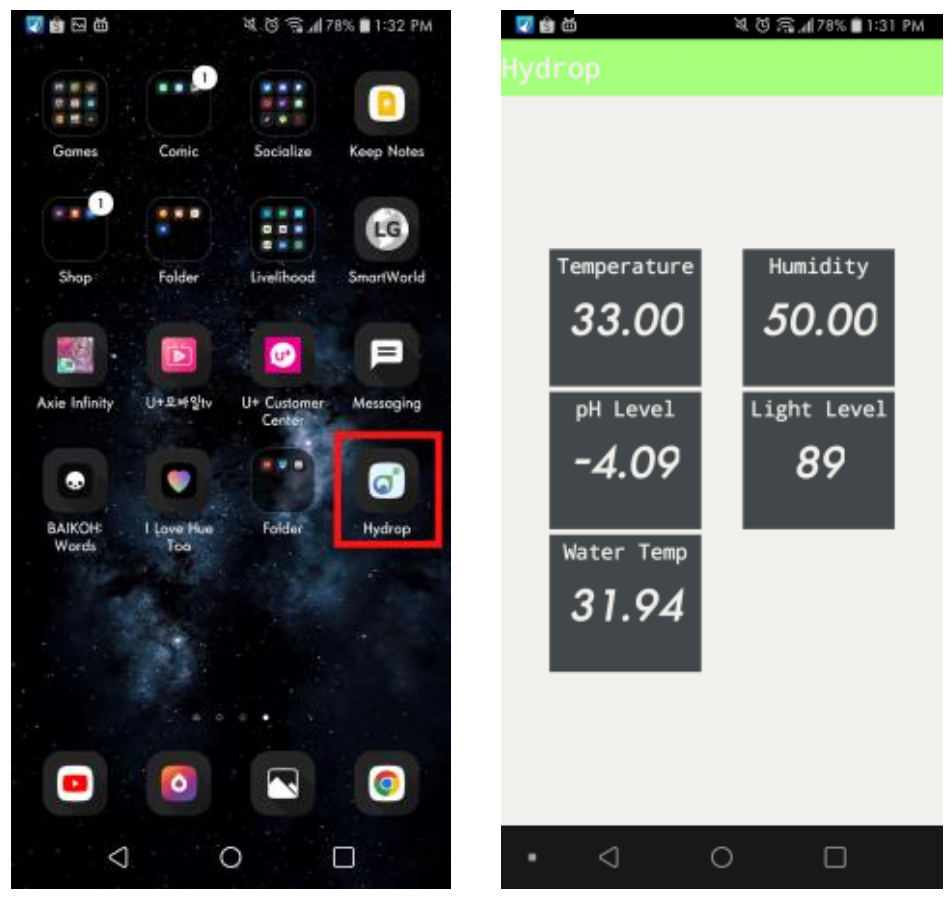

*Figure 10.c*. Developed Android Mobile Application for SMART Greenhouse

Figure 10.a – 10.c shows the components and stages of the developed mobile application used to monitor the hydroponic garden within the greenhouse setting. The developed mobile application used ThinkSpeak and MIT App Inventor software to achieve the goal of the study.

Shown in Table 2-6 were the actual sample data logs for monitoring the temperature, humidity, pH level, light level, and water temperature level of the SMART greenhouse which was conducted by the researchers during the testing stage. Below is the explanation of the data that has been displayed during the testing.

Table 2. Temperature data from ThinkSpeak

| Date | Time | Data |
|---|---|---|
| 05-23-2022 | 16:19 | 26C |
| 05-23-2022 | 16:26 | 26C |
| 05-23-2022 | 19:08 | 29C |
| 05-24-2022 | 17:27 | 28C |
| 05-24-2022 | 17:48 | 28C |
| 05-24-2022 | 18:10 | 27C |
| 05-25-2022 | 10:31 | 34C |
| 05-25-2022 | 10:38 | 32C |
| 05-25-2022 | 11:01 | 33C |

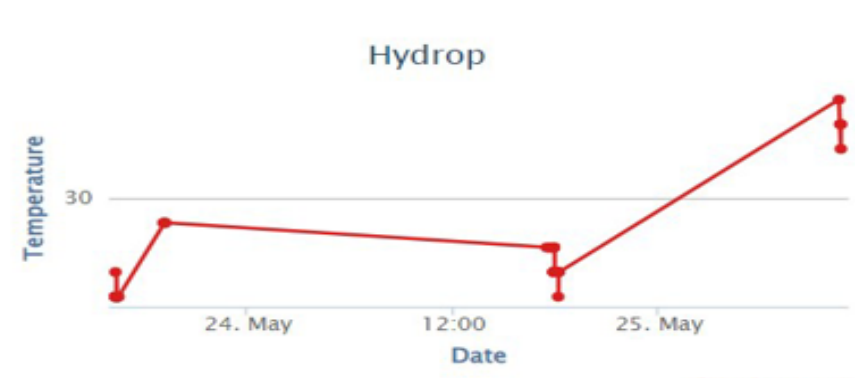

Table 2 shows that the highest temperature that is recorded at the time of testing is 34C which happened on May 25, 2022, at 10:31 AM. The temperature rise was due to the weather being sunny at that time. While the lowest temperature recorded was 26C, which is on May 23, 2022, at 4:19 PM, and during that time it is partly cloudy and had light rain.

Table 3. Humidity data from ThinkSpeak

| Date | Time | Data |
|---|---|---|
| 05-23-2022 | 16:19 | 83% |
| 05-23-2022 | 16:26 | 81% |
| 05-23-2022 | 19:08 | 77% |
| 05-24-2022 | 17:27 | 77% |
| 05-24-2022 | 17:48 | 84% |
| 05-24-2022 | 18:10 | 73% |
| 05-25-2022 | 10:31 | 46% |
| 05-25-2022 | 10:38 | 48% |
| 05-25-2022 | 11:01 | 50% |

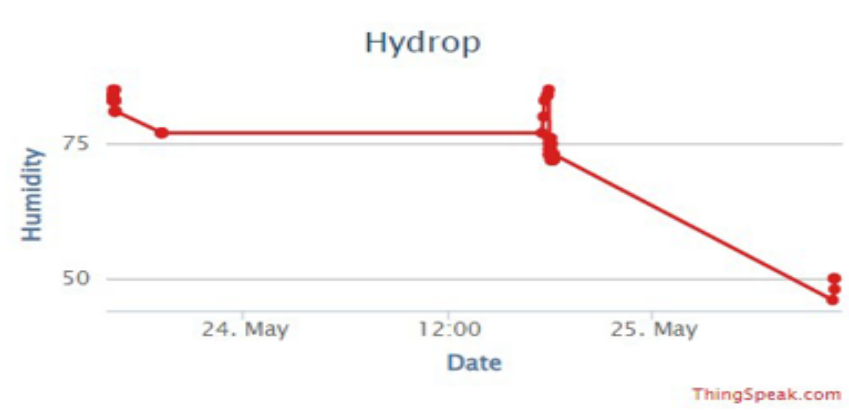

Table 3 shows the humidity sample logs during the testing period of the prototype. At 5:48 PM on May 24, 2022, the highest peak of humidity (84%) was recorded and on May 25, 2022, the lowest (46%) was recorded. During that time the humidity of the place depends on the weather that day. That is the reason why those two times are different.

Table 4. pH Level data from ThinkSpeak

| Date | Time | Data |
|---|---|---|
| 05-23-2022 | 16:19 | 2.96 |
| 05-23-2022 | 16:26 | 2.74 |
| 05-23-2022 | 19:08 | 0.58 |
| 05-24-2022 | 17:27 | 6.61 |
| 05-24-2022 | 17:48 | 7.01 |
| 05-24-2022 | 18:10 | 7.19 |
| 05-25-2022 | 10:31 | -2.54 |
| 05-25-2022 | 10:38 | -4.09 |
| 05-25-2022 | 11:01 | -3.89 |

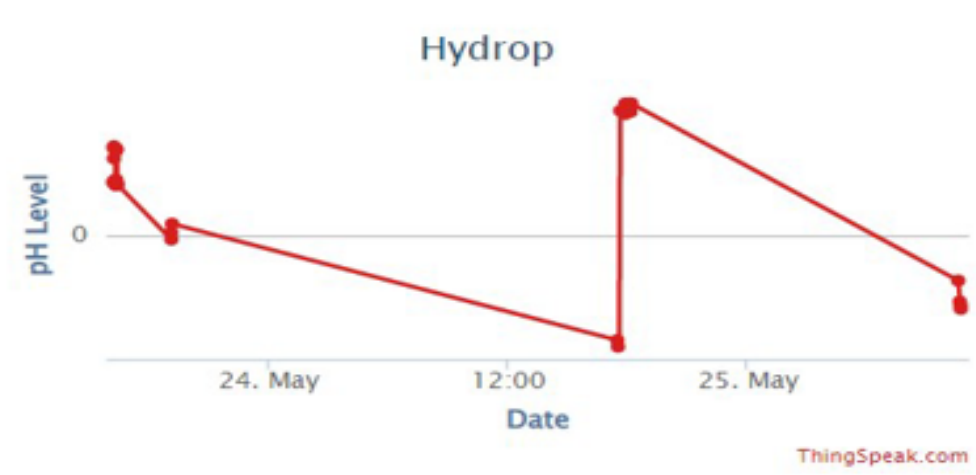

Table 4 shows the pH Level recorded on the days of trials. The highest recorded data is 7.19 and the lowest being -4.09. The negative result of the trial was caused by the improper calibration of the pH sensor during testing. pH sensor calibration is required to correctly identify the pH level of the water.

Table 5.  Light Level data from ThinkSpeak

| Date | Time | Data |
|---|---|---|
| 05-23-2022 | 16:19 | 384 |
| 05-23-2022 | 16:26 | 391 |
| 05-23-2022 | 19:08 | 1013 |
| 05-24-2022 | 17:27 | 82 |
| 05-24-2022 | 17:48 | 83 |
| 05-24-2022 | 18:10 | 80 |
| 05-25-2022 | 10:31 | 83 |
| 05-25-2022 | 10:38 | 84 |
| 05-25-2022 | 11:01 | 89 |

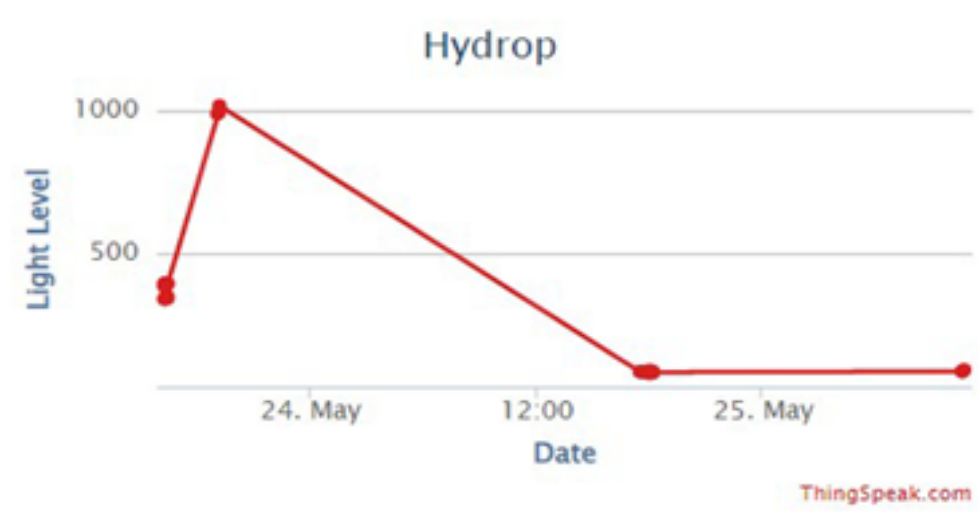

Table 5 shows the light level of the sample logs. The highest that was recorded is 1013 on May 23, 2022, at 7:08 PM and the lowest is 80 on May 24, 2022, at 6:10 PM. The weather is also one of the reasons that affect the light level of the greenhouse. Another factor is the time that the data is taken and the light in the surrounding area of the greenhouse.

Table 6.  Water Temperature data from ThinkSpeak

| Date | Time | Data |
|---|---|---|
| 05-23-2022 | 16:19 | 26.63 C |
| 05-23-2022 | 16:26 | 26.06 C |
| 05-23-2022 | 19:08 | 29.5 C |
| 05-24-2022 | 17:27 | 28.75 C |
| 05-24-2022 | 17:48 | 28.31 C |
| 05-24-2022 | 18:10 | 27.63 C |
| 05-25-2022 | 10:31 | 32.69 C |
| 05-25-2022 | 10:38 | 31.75 C |
| 05-25-2022 | 11:01 | 31.94 C |

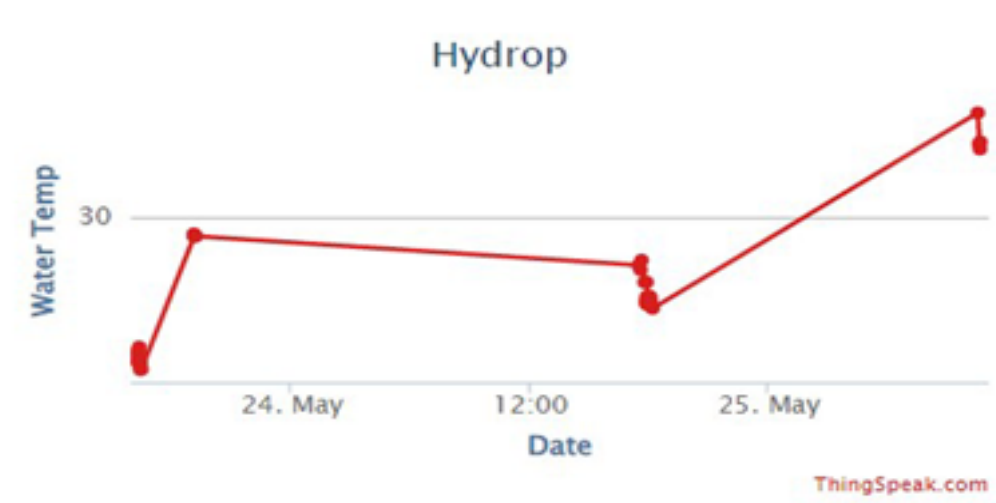

On May 25, 2022, at 10:31 AM, the highest reading was performed with a reading of 32.69 C and on May 23, 2022, at 4:26 PM the lowest reading happened (26.06 C). The change in temperature occurred because of the climate at that time. Additionally, the temperature inside the greenhouse also affects the temperature of the water.

The outcome of the testing as displayed in Table 2 – 6 shows that the results data is within range of the parameter required for adequate growth of the hydroponic plants which were for pH is 6.5 to 8, temperature of greenhouse is 26C to 29C, temperature of water is 28C to 31C, humidity is 70% where trials were done once a day.

## *Results of Trial Testing of the Prototype in Comparison to Commercialized Device*

To validate the outcome of the study, the proponents conducted both prototype and commercialized device series of tests and take notes on the comparison of the results. The computed percentage difference between the test results of the project prototype and the commercialized device was used to verify the success of the study.

Table 7 displays the results of the testing done by the proponents concerning the water pH level and temperature, greenhouse temperature, humidity, and light. The data gathered from the prototype and the commercialized device were not as far from each other as shown in the table. When it comes to the pH level as shown in the table, the results fall within the required range, which is 6.8 to 8 as stated in the study by (Judith, 2019). In the test trials done for water temperatures, the results were close to each other and meet the required range of 28°C to 31°C according to (Robles, 2022). The data presented in Table 7 regarding the greenhouse temperature, the results are in accordance with the range of 26°C to 29°C and with a percentage difference not being greater than 3.5% shows the precision of the prototype according to (Robles, 2022). In terms of humidity, the required basis for good plants growth must not be lower than 70% (AdvancedNutrients.com, 2017), and compared with the prototype results with 6.8% percentage difference, it means that the sensor used in the prototype is not reliable compared to the commercialized device. For light trial results seen in Table 7 with having a percentage difference not exceeding 2.4%, it can be said that the data gathered from the sensors are precise. With the use of the percentage difference formula, the data can prove the reliability of the sensor in the prototype.

Illustrated in Table 8 shows the different sensors that are included in the greenhouse system. The table shows the 3 trials that were conducted to determine if the sensors are gathering data that is within the standard value range. The table shows that almost all the sensors read a value within the range; however, during trial 3 the PH4502C sensor reads a negative value that is not within the range. This is due to the reason that the calibration of the PH4502C was not working properly at that time. Overall, this table shows that the sensors read the data well. Hence, the Sensor system is working successfully.

Table 7. Project Prototype and Commercialized Device Testing Results

| | Protype | | | Commercialized Device | | | Percentage Difference | | |
|---|---|---|---|---|---|---|---|---|---|
| | Trial 1 | Trial 2 | Trial 3 | Trial 1 | Trial 2 | Trial 3 | Trial 1 | Trial 2 | Trial 3 |
| pH Sensor (pH4502c with pH Probe) | 7.18 | 7.77 | 7.77 | 7.73 | 7.7 | 7.74 | 7.37% | 0.9% | 0.3% |
| Water Temperature Sensor (Probe) | 28.88° C | 27.75 °C | 27.94 °C | 28.5°C | 27.8°C | 28.6°C | 1.32% | 0.18% | 2.33% |
| Greenhouse Temperature (DHT-22) | 28 °C | 28.9 °C | 27 °C | 29°C | 28°C | 27°C | 3.5% | 3.1% | 0% |
| Humidity (DHT- 22) | 73% | 76% | 75% | 70% | 71% | 70% | 4.1% | 6.8% | 6.8% |
| Light (LDR Module) | 83% | 82% | 83% | 84% | 80% | 81% | 1.1% | 2.4% | 2.4% |

Note: Percentage Difference: $\frac{|V1-V2|}{\frac{(V1+V2)}{2}}\ x\ 100$

Table 8. Sensor Data Gathering

| Sensor | Standard Value Range | Trial 1 | Trial 2 | Trial 3 | Remarks |
|---|---|---|---|---|---|
| DHT11 (Temperature) | 0 °C - 50 °C | ✓ | ✓ | ✓ | All are within range |
| DHT11 (Humidity) | 20% - 90% | ✓ | ✓ | ✓ | All are within range |
| PH4502C | 0 – 14 | ✓ | ✓ | ✗ | 2 out of 3 are within range |
| DS18B20 | -55 °C - + 125 °C | ✓ | ✓ | ✓ | All are within range |
| LDR Module | 0 – 1023 | ✓ | ✓ | ✓ | All are within range |

## *Evaluation Results*

To evaluate the prototype's functionality, reliability, and usability, the proponents chose a total of ten (10) respondents, as it is enough to build a basis for reliable evaluation results (Graglia, 2022). Through the convenience sampling technique, the survey uses a five-point Likert scale as advised with similar questionnaires (Pollfish, 2022).

The responses to this survey indicate the suitability of the system when it comes to being functional, reliable, and usable, therefore using the ISO 9126 survey as an adaptation for the system's questionnaire. The questionnaire for the mobile application was adapted from MARS by the Queensland University of Technology, the mobile application rating scale, and a few questions were solely made by the proponents.

Table 9 shows the result of the evaluation regarding the system's functionality. Based on the data gathered as rated using the Likert scale, the respondents agreed that the system functions according to its intended purpose.

Table 9. Functionality - Descriptive Statistics Evaluation Output of the Prototype

| Criteria | Mean | Std. Deviation | Interpretation |
|---|---|---|---|
| 1. The system and its components are functioning properly in accordance with their intended purpose. | 4.30 | 0.675 | Strongly Agree |
| 2. The system provides accurate information in the database. | 4.30 | 0.483 | Strongly Agree |
| 3. The tool works well with the computer system that is currently in use. | 4.20 | 0.789 | Strongly Agree |
| 4. The technology can test the physical environment in its entirety. | 4.10 | 0.738 | Agree |
| 5. The inputs are accurately represented by the system. | 3.70 | 0.949 | Agree |
| 6. The system can assist with managerial responsibilities. | 4.30 | 0.675 | Strongly Agree |
| 7. The system is capable of handling complete project testing. | 3.90 | 0.876 | Agree |
| 8. The system can operate for an extended amount of time without crashing. | 4.00 | 0.816 | Agree |
| 9. The system complies with the testing methodology's requirements. | 4.30 | 0.483 | Strongly Agree |
| Grand Mean | 4.12 | 0.365 | Agree |
| Functionality Verbal Equivalent | Very Good | | |

Note: For interpretation, the following remarks apply to mean interval: 5.00 – 4.20 for Strongly Agree (Excellent), 4.19 – 3.40 for Agree (Very Good), 3.39 – 2.60 for Slightly Agree (Good), 2.59 – 1.80 for Disagree (Fair), and 1.79 – 1.00 for Strongly Disagree (Deficient).

Table 10 displays the results from respondents concerning the system's reliability. The interpretation ranges from agreeing to strongly agreeing at best, declaring that the system is reliable when used correctly.

Table 10. Reliability - Descriptive Statistics Evaluation Output of the Prototype

| A | Mean | Std. Deviation | Interpretation |
|---|---|---|---|
| 1. The system is capable of repeating its task throughout time. | 4.20 | 0.632 | Strongly Agree |
| 2. When the internet is available, the system connects to it. | 4.40 | 0.699 | Strongly Agree |
| 3. In the event of a failure, the system is capable of recovering data. | 3.90 | 0.738 | Agree |
| 4. After prolonged use, the system does not crash. | 4.10 | 0.568 | Agree |
| 5. On the device in use, the system works perfectly. | 4.40 | 0.699 | Strongly Agree |
| Grand Mean | 4.20 | 0.298 | Strongly Agree |
| Reliability Verbal Equivalent | Excellent | | |

Table 11 shows the results in which the system's usability was evaluated. Overall, according to the system's uses, the respondents agree that the system is simple and easily understandable enough to be used by them.

Table 11. Usability Descriptive Statistics Evaluation Output of the Prototype

| Criteria | Mean | Std. Deviation | Interpretation |
|---|---|---|---|
| 1. Overall, the technique is simple to comprehend. | 4.40 | 0.516 | Strongly Agree |
| 2. The system's functions are simple to use. | 4.70 | 0.483 | Strongly Agree |
| 3. The system's functions are easily accessible and well-organized. | 4.10 | 0.568 | Agree |
| 4. The system is simple to use and understand. | 4.00 | 0.816 | Agree |
| 5. The system provides users with appropriate guidance. | 3.80 | 0.789 | Agree |
| 6. The system's user interface is well-designed. | 4.00 | 0.816 | Agree |
| Grand Mean | 4.17 | 0.416 | Agree |
| Usability Verbal Equivalent | Very Good | | |

Table 12 made use of MARS or Mobile Application Rating Scale, which tested the different aspects of the mobile application when downloaded and in use. In terms of engagement, the respondents agreed that the mobile application was engaging for them when in use.

Table 12. Mobile Application Descriptive Statistics Evaluation Output

| Criteria for Mobile Application using MARS | Mean | Std. Deviation | Interpretation |
|---|---|---|---|
| 1. *Interest* - These employs ways to boost engagement by presenting material in an engaging manner. | 3.80 | 0.422 | Agree |
| 2. *Customization* -The app provides/retains all relevant app settings/preferences (e.g., sound, content, notifications, etc.) | 3.80 | 0.919 | Agree |
| 3. *Interactivity* - This enables for user interaction, feedback, and prompts (reminders, sharing options, notifications, etc.) | 3.80 | 1.033 | Agree |
| 4. *Target Group* - The content of the app (visual data, language, and design) is acceptable for the intended audience. | 4.10 | 0.316 | Agree |
| Grand Mean | 3.88 | 0.445 | Agree |
| Engagement Verbal Equivalent | | Very Good | |

Table 13 evaluated the mobile application's functionality, and how it links into the system at hand. Based on the results, the respondents were happy with the level of functions that lies within the application.

Table 13. Mobile Application Descriptive Statistics Evaluation Output by Functionality

| Criteria for Mobile Application using MARS | Mean | Std. Deviation | Interpretation |
|---|---|---|---|
| 1. *Performance* - The app's features (functions) and components (buttons/menus) all perform correctly and quickly. | 3.80 | 0.919 | Agree |
| 2. *Performance* - Does accessing other tab of the application is easy to understand? | 4.30 | 0.675 | Strongly Agree |
| 3. *Performance* - Does the application give correct data? | 3.80 | 0.789 | Agree |
| 4. *Performance* - Does every button of mobile application provide the correct output? | 3.80 | 0.789 | Agree |
| 5. *Ease of Use* - The menu labels/icons are simple to understand and use. | 4.00 | 0.667 | Agree |

Table 13. Mobile Application Descriptive Statistics Evaluation Output by Functionality (cont.)

| Criteria for Mobile Application using MARS | Mean | Std. Deviation | Interpretation |
|---|---|---|---|
| 6. *Ease of Use* - The app's instructions are simple and straightforward. | 4.20 | 0.789 | Strongly Agree |
| 7. *Ease of Use* - All screen linkages offered are present; moving between displays is logical/accurate/appropriate/uninterrupted. | 4.30 | 0.483 | Strongly Agree |
| 8. *Ease of Use* - All components/screens have consistent and intuitive interactions (taps/swipes/pinches/scrolls). | 3.80 | 0.632 | Agree |
| Grand Mean | 4.00 | 0.312 | Agree |

Table 14 describes the response of the people who evaluated the mobile application in terms of its aesthetics or UI. The consensus was ranging from agree to strongly agree, therefore the application UI was adequate and usable for the system utilized.

Table 14. Mobile Application Descriptive Statistics Evaluation Output by Aesthetics

| | Mean | Std. Deviation | Interpretation |
|---|---|---|---|
| 1. *Layout* - The placement and size of buttons, icons, menus, and information on the screen suitable and they can be zoomed if necessary. | 3.60 | 0.516 | Agree |
| 2. *Layout* - The graphical user interface is easy to understand and use. | 4.10 | 0.568 | Agree |
| 3. *Graphics* - The visuals used for buttons, icons, menus, and content are of excellent quality and resolution. | 3.90 | 0.738 | Agree |
| 4. *Visual Appeal* - The software appears to be of high quality. | 3.40 | 0.699 | Agree |
| 5. *Visual Appeal* - The system is beneficial to the users. | 4.20 | 0.632 | Strongly Agree |
| 6. *Visual Appeal* - The prototype is comfortable and easy to use. | 4.10 | 0.876 | Agree |
| Grand Mean | 3.88 | 0.360 | Agree |
| Aesthetic Verbal Equivalent | Very Good | | |

Lastly, table 15 evaluates the reliability of the information seen in the application. The respondents agree that all the results they see on display were accurate, and very reliable.

Table 15. Mobile Application Descriptive Statistics Evaluation Output by Information

| Criteria for Mobile Application using MARS | Mean | Std. Deviation | Interpretation |
|---|---|---|---|
| 1. *Accuracy of App Description* - The app contains features and functions correctly based on its descriptions. | 4.60 | 0.516 | Strongly Agree |
| 2. *Goals* - Are there explicit, quantifiable, and achievable goals in the app (described in the app store description or within the app)? | 3.60 | 0.843 | Agree |
| 3. *Quality of Information* - The material is accurate, well-written, and pertinent to the app's goal/topic. | 4.10 | 0.316 | Agree |
| 4. *Quantity of Information* - The information coverage is thorough but succinct and is within the scope of the app. | 4.10 | 0.738 | Agree |
| 5. *Visual Information* - Concepts are visually explained in a clear, logical, and proper manner, using charts, graphs, photos, videos, and other visual aids. | 3.90 | 0.738 | Agree |
| Grand Mean | 4.06 | 0.212 | Agree |
| Information Verbal Equivalent | | Very Good | |

Table 16 shows the Cronbach's Alpha result regarding the reliability of the questionnaire given by the proponents to the respondents. With the given data, the survey questions had 76.40% reliability.

Table 16. Reliability Statistics Output of System Evaluation

| Cronbach's Alpha | Cronbach's Alpha Based on Standardized Items | N of Items |
|---|---|---|
| 0.764* | 0.726 | 20 |

Note: * - The statements under the system evaluation obtained 76.40% reliability.

Table 17 states the mobile UI evaluation questions were 92.40% reliable. This goes to show that both survey questionnaires done by the proponents are reliable when given to and used by the respondents.

Table 17. Reliability Statistics Output on the Mobile UI Evaluation

| Cronbach's Alpha | Cronbach's Alpha Based on Standardized Items | N of Items |
|---|---|---|
| 0.924* | 0.920 | 23 |

Note: * - The statements under the mobile UI evaluation obtained 92.40% reliability.

Table 18 shows the results of the trials in comparison, which were done to get reliable data from the system prototype and the commercialized devices bought by the proponents. Close margins were observed based on the results and will be discussed further in this part of the document.

Table 18. Data Entry of the Parameters Generated by Prototype and Commercialized Device on the Reliability Test Results

| Trial | pH Level | | Temperature of the Greenhouse (in °C) | | Temperature of the Water (in °C) | | Humidity | | Light | |
|---|---|---|---|---|---|---|---|---|---|---|
| | P | C | P | C | P | C | P | C | P | C |
| 1 | 7.18 | 7.73 | 28 | 29 | 28.88 | 28.5 | 73 | 70 | 83 | 84 |
| 2 | 7.77 | 7.70 | 28.9 | 28 | 27.75 | 27.8 | 76 | 71 | 82 | 80 |
| 3 | 7.77 | 7.74 | 27 | 27 | 27.94 | 28.6 | 75 | 70 | 83 | 81 |

Note: P – indicates the data generated by the Prototype. C – indicates the data generated by commercialized Device. The ideal range for pH is 6.5 to 8, temperature of the greenhouse is 260C to 290C, temperature of water is 280C to 310C, humidity is 70%. The trials were done once a day.

In this part of the paper, testing happened every day for three days to garner such results. Tables in part display the data gathered from those days of testing. The testing phase also garnered each parameter needed for this study, mainly the pH level, the greenhouse humidity, and temperature, as well as light, and the water temperature. According to the results of the tests, the data acquired shared a resemblance to the ideal measurements of parameters needed to grow a functioning hydroponic system. Accordingly, the tables above present the results of the functionality, reliability, and usability testing of this study. Alpha tests were done to know if the prototype is working to the extent needed by the proponents, to know the expected output of the prototype, and check the process of the components utilized.

Table 10-18 displayed the evaluation results from the project`s outcome in terms of functionality, reliability, and usability with highly commendable results performed by the selected respondents of the study. Based on data gathered from the series of tests and evaluation above, precision between the system's sensors data and the commercialized

data are seen appropriately, and the system and its mobile application evaluated can be concluded as highly reliable.

## CONCLUSIONS AND RECOMMENDATIONS

Nowadays, a lot of people wanted to grow plants at their own expense but have no way of doing so due to lack of space and time for monitoring since most are out working during. To aid this concern, the proponents successfully designed and developed an IoT-Based Controlling and Monitoring Apparatus on a Greenhouse for hydroponic gardens, as well as designed and developed a controlling and monitoring Android Mobile Application for the greenhouse system using MIT App Inventor.

The researchers managed to construct an IoT-Based Greenhouse Hydroponic system that can monitor the pH level, Light, water and greenhouse Temperature, and Humidity. An IoT-Based Greenhouse Hydroponic system that manipulates pH level and lowers the water temperature was also developed successfully and was tested and evaluated using ISO 9126, in terms of its functionality, reliability, and usability.

The proponents discovered the needs and requirements of different plants such as lettuce, kale, and basil when grown hydroponically and grown traditionally. With the use of IoT based platforms such as ThingSpeak, the convenience of automated systems such as the IoT Smart Greenhouse System is successfully defined in this study. The IoT Smart Greenhouse System for Hydroponic Gardens can be concluded to be an effective system to grow and monitor plants effectively. Without the use of excessive space and soil and any sort of pesticides, this study has proven that monitoring and growing plants can be possible just from the application itself.

As for recommendation, this system can be improve by the future researchers in the following aspects: (1) the use of Solar Panels, so that the pump will not be relying on the AC power alone; (2) the wiring system of the prototype should be remodeled; (3) the usage of bigger microcontroller would be advisable to be able to accommodate more sensors and devices for a larger arsenal of data to be gathered; (4) adopt higher Arduino microcontroller model for this kind of project; (5) updates such as bug fixes for the application are also needed, as well as having a note on the application that shows the required ranges of the parameters in educate new users; (6) enclosures of the devices to ensure the safety of the sensors; (7) the proponents also recommend a longer period of testing, to get data that can further strengthen the reliability of the results; and (7) the ventilation will be controlled automatically when the ideal temperature goes above or below the set data.

## ACKNOWLEDGEMENT

The proponents would like to express their deepest gratitude to all those who had been part of the completion of this study. First, to their thesis adviser, Engr. Jocelyn

Bernardino, for her persistent support and active participation in the research paper writing process at every level, and to Engr. Emmanuel O. Robles, for his guidance and support to the proponents as they constructed their greenhouse. They would also like to thank Dr. Arnel M. Avelino, Dean - College of Engineering, Computer Studies, and Architecture, the Research and Knowledge Management Department, and the administration of LPU-Cavite for their utmost support and inspiration throughout the study.

## DECLARATIONS

### *Conflict of Interest*

The authors declares that there is no conflict of interest.

### *Informed Consent*

The data used in this study was collected from developed prototype thru a series of testing and to the target respondents with consent and well informed to its purpose. The participant's identity was not revealed and not included in the documentation.

### *Ethics Approval*

The LPU-Cavite Center for Research and Knowledge Management and the Research Ethics Committee accepted and approved the conduct of the study.

## Author's Biography

Arcel Christian H. Austria was born on December 10, 1999, in Cavite City. He went to Lyceum of the Philippines University to take up Bachelor of Science in Computer Engineering in Gen. Trias Cavite from 2018 to 2022. He participated in TESDA programs earning him the NCII Certificate. He is a former president and a member of Institute of Computer Engineers of the Philippines (ICpEP).

John Simon Fabros was born on August 12, 2000, in Indang, Cavite. He obtained his bachelor`s degree in Computer Engineering at Lyceum of the Philippines University - Cavite. He attended numerous seminars of the organization to further enhance his knowledge and skills, and a member of The Institute of Computer Engineers in the Philippines (ICPEP).

Kurt Russel G. Sumilang was born on November 20, 2000 in Cavite. He went to Lyceum of the Philippines to take up Bachelor of Science in Mechanical Engineering in Gen. Trias Cavite from 2018 to 2022. He attended numerous seminars of the organization to further enhance his knowledge and a member of The Institute of Computer Engineers in the Philippines (ICPEP).

Jocelyn B. Bernardino is a college instructor at the College of Engineering, Computer Studies, and Architecture (COECSA) and the Head of COECSA Laboratories of Lyceum of the Philippines University – Cavite (LPUC). She holds a B.S. degree in Electronics and Communication Engineering (BSECE) and ongoing MEng'g in Computer Engineering from Pamantasan ng Lungsod ng Maynila (PLM). She is a member of Institute of Computer Engineers of the Philippines (ICpEP), and a recipient of database fundamentals certification. She co-authors technical paper of undergraduate engineering students of LPUC and her works focuses on microcontroller and IOT applications.

Anabella C. Doctor is a Computer Engineering Department chair at Lyceum of the Philippines University – Cavite. She finished several research focused on software development and related fields, presented in both local and international conferences with journal publication which was funded by a university and Commission on Higher Education. She finished Master's in Information Technology at TUP, Manila and currently taking Doctor of Information Technology at DLSU - Dasmariñas. She is proficient web programming, Visual C# and Visual Basic programming language with MS SQL for database. She is also an author of two computer books used in junior high school.